\documentclass{article}
\usepackage{graphicx}
\usepackage{float}
\usepackage{amssymb}
\newcommand{\be}{\begin{equation}}
\newcommand{\ee}{\end{equation}}
\begin{document}

\title{Modeling power grids}

\author{Per Arne Rikvold, Ibrahim Abou Hamad, Brett Israels\\
Department of Physics,
Florida State University\\
Tallahassee, FL 32306-4350, U.S.A.\\
{~}\\
Svetlana V.\ Poroseva\\
Mechanical Engineering Department, University of New Mexico\\  
Albuquerque, NM 87131-0001, U.S.A.}

\date{\today}

\maketitle

\section*{Abstract}
We present a method to construct random model power grids that closely match statistical properties of a real power grid. The model grids are more difficult to partition than a real grid. 


\section{Introduction}
\label{sec-Int}
Power grids are prime examples of the  interconnected networks that make life in technological societies possible. It is therefore of paramount importance to develop methods to prevent cascading failures that manifest themselves in widespread, catastrophic blackouts. One such defensive strategy is Intentional Intelligent Islanding  \cite{LILI05,PEIR09}, which aims to prepare for the rapid isolation of parts of the grid where instabilities arise before they can spread further. This problem of network partitioning \cite{NEWM04C,FORT10} involves determining subdivisions of the grid that are tightly internally connected, but weakly connected to the rest of the grid. At the same time, each such island should be close to self-sufficient with power. 

In previous papers \cite{HAMAD2010,HAMAD2011} we have considered network-theoretical methods to achieve this goal, using the high-voltage network in the U.S.\ state of Florida as a test example. Here we continue this endeavor by constructing model power grids that share important statistical properties with a real one. Such models provide opportunities to study the effects of specific network modifications, as well as to perform scaling analyses in terms of the network size. 

\section{Model construction}
\label{sec-Mod}
The Florida high-voltage power grid~\cite{FLAMAP} consists of $N=84$ vertices, 31 of which are generators and the rest are distribution substations that act as  loads. The vertices are connected by $M=200$ edges, some of which are parallel power lines connecting the same two vertices. 
We use a simplified representation of the grid as a weighted, undirected graph \cite{NEWM04C}, defined by the 
$N \times N$ symmetric weight matrix $\bf W$, whose
elements $w_{ij} \ge 0$ represent the ``conductances'' of 
the edges (transmission lines) between vertices (generators or loads)
$i$ and $j$,
\begin{equation}
\label{eq:conductivities}
w_{ij}=\frac{ {\rm number\, of\, lines\, between\, vertices\,}i\,{\rm and}\,j}{\rm{\, normalized
\, geographical \, distance}}, 
\end{equation}
where the ``geographical distance'' is the length of the edge connecting 
$i$ and $j$. To obtain a model independent of any specific systems of length units, the distances are normalized such that the areal density of vertices is unity. In Fig.~\ref{fig-FL} we show a map of the Florida network together with a representative model network. 

\begin{figure}
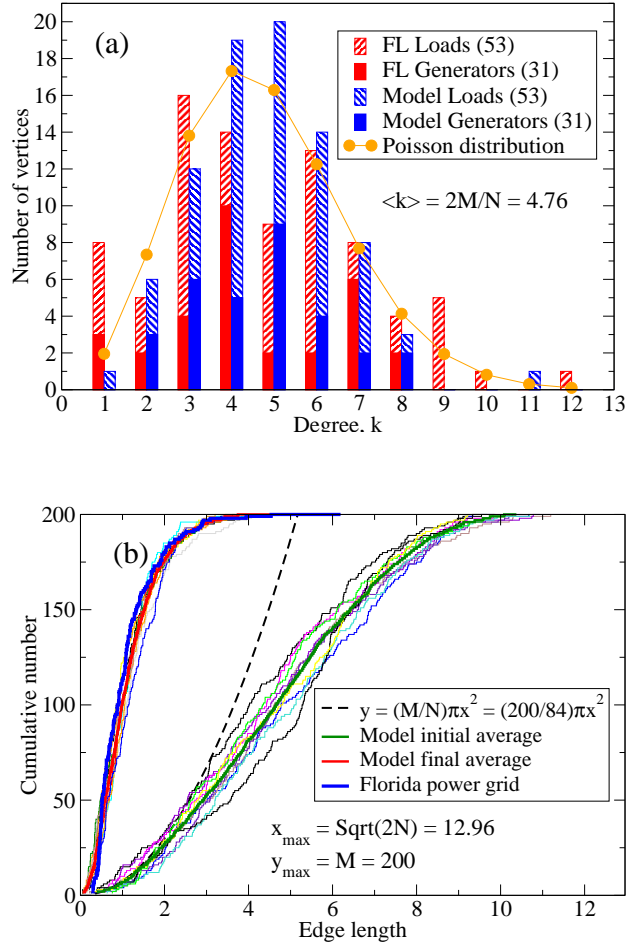

\begin{center}
\includegraphics[angle=0,height=0.48\textwidth]{HistFL_RFL6_GenLoads.eps}
\vspace{0.5truecm}
\end{center}
\begin{center}
\includegraphics[angle=0,height=0.48\textwidth]{DistDistUGA12.eps} 
\end{center}
\caption[]{
(a) Degree distributions for the Florida power grid (left set of bars, in red) 
and the representative model grid analyzed here (right set of bars, in blue). 
Also shown is a Poisson distribution for $k-1$ with mean $2M/N - 1$. 
(b) Cumulative edge-length distributions, normalized to $M=200$. Ten 
independent realizations of the model are included (thin curves in 
the background), together with their average distributions (heavy 
curves in the foreground). The right-hand set of curves represents the 
initial, ``infinite-temperature" distributions (their average gives an 
estimate for the model's density of states). The left-hand set of curves 
represents the final, ``low-temperature" distributions at 
$T = 1.09/2 = 0.545$. The bold, blue curve on the left represents the 
distribution for the real Florida grid. Its average edge length is 
approximately 1.09.  The dashed parabola represents the theoretical  model 
density of states for an infinite system with unit areal vertex density and 
average degree $2M/N$. It fits the simulation well for short edge lengths.
}
\label{fig-distr}
\end{figure}

\begin{figure}
\begin{center}
\includegraphics[angle=0,height=0.55\textwidth]{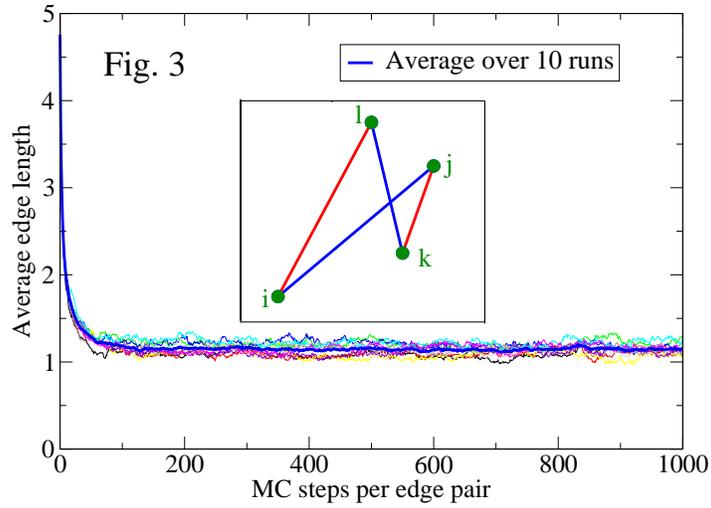} 
\end{center}
\caption[]{
Average edge length vs.\ MC steps per edge. Ten independent simulation runs (thin curves in the background) and their average (thick curve in the foreground). Inset: schematic of the MC update $ij,kl$ (blue) $\rightarrow il,kj$ (red).
}
\label{fig-cool}
\end{figure}

Model networks were produced by the following procedure.
\begin{enumerate}
\item
We placed the $N=84$ vertices randomly in a square of area $N$.
\item
Following the standard ``stub" method \cite{NEWM10}, we attached $2M=400$ stubs or half-edges randomly to the $N$ vertices. (Actually, to ensure that no vertices in this small network should be totally isolated, we first attached one stub to each vertex, and then distributed the remaining $2M-N$ stubs randomly between the vertices.) The resulting degree distribution for the particular model grid discussed in this paper 
is shown together with that of the real Florida grid in Fig.~\ref{fig-distr}(a). 
\item
We connected the stubs randomly in pairs, with the restriction that self-loops (two mutually connected stubs at the same vertex) were forbidden. 
\item
To obtain an edge-length distribution with the same average as that of the real Florida grid ($\approx 1.09$ in our dimensionless units), we employed a Monte Carlo (MC) ``cooling" procedure using a ``Hamiltonian" in which the total edge length $L$ plays the role of the system energy. The update attempts consisted in choosing two different edges, $ij$ and $kl$ with $i \neq j \neq k \neq l$, interchanging $j$ and $l$, and calculating the change $\Delta L = [L(ij)+L(kl)]-[L(il)+L(kj)]$. Attempts were accepted with the Metropolis probability with a fictitious ``temperature" $T$, $P(ij,kl \rightarrow il,kj) = {\rm Min}[1,\exp(-\Delta L /T)]$. In the limit that all edges are $\ll \sqrt{N}$, it is easy to show that the partition function for this model is $Z = 2 \pi T^2$, and so $\langle L(ij) \rangle = 2T$. 
Edge length distributions before and after ``cooling" are shown  together with that of the real Florida grid in 
Fig.~\ref{fig-distr}(b). The average edge length vs.\ the number of MC steps is shown in Fig.~\ref{fig-cool}, together with a schematic of the MC update mechanism. 
\end{enumerate}

\section{Network partitioning}
\label{sec:met}
The method used to partition the grids was described in Ref.~\cite{HAMAD2011} and is only sketched here.

\noindent
{\bf Agglomeration.} 
A trial partitioning is obtained by associating each load $i$ with its ``nearest" generator $j$, defined 
as the one to which the {\it effective resistance\/} \cite{KLEIN93}, ${\bf R}_{ij}$ is minimum. 

\noindent
{\bf Optimization.}
The goal is to obtain a partitioning into $K$ islands, ${\mathcal C} = \{C_1, ..., C_K\}$, that balances the requirements for islands that are (i) strongly connected internally, but sparsely connected to each other, and (ii) approximately self-sufficient with power. 

Success in achieving requirement (i) is measured by the 
{\it modularity\/} $Q$ \cite{NEWM04C}. It compares the proportion of edges 
internal to islands with the same proportion in an
average null-model. 
\be
Q = \frac{1}{w} \sum_{ij} \left( w_{ij} - \frac{w_i w_j}{w}  \right)
\delta \left( C(i),C(j) \right)
\;,
\label{eq:Q}
\ee
where $w_i = \sum_j w_{ij}$, $w = \sum_{ij} w_{ij}$, and 
$\delta \left( C(i),C(j) \right) = 1$ if vertices $i$ and $j$ belong to the
same island, and $0$ otherwise. 
(Other quality measures could also be used \cite{RONH2010}.)

\begin{figure}
\vspace{1.0truecm}
\begin{center}
\includegraphics[angle=0,height=0.55\textwidth]{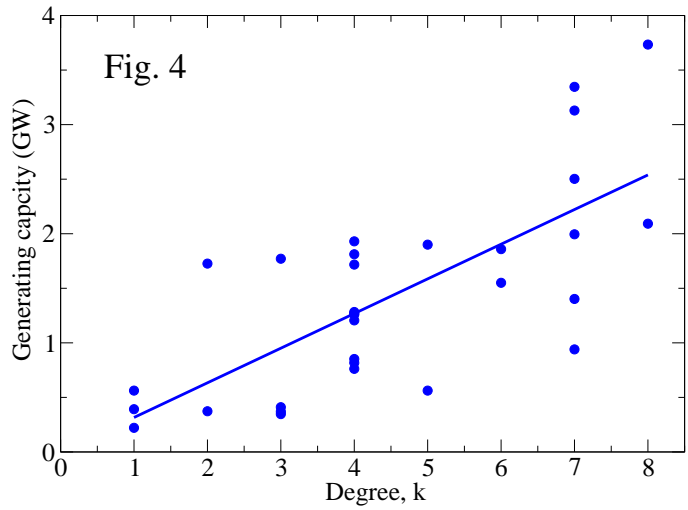} 
\end{center}
\caption[]{
Generating capacities of Florida power plants vs.\ their degrees. The line is a least-squares fit to the data.
}
\label{fig-capdeg}
\end{figure}

We show in Ref.~\cite{HAMAD2011} that success in achieving requirement (ii) requires minimizing the sum of the squares of the currents $I_i$ that enter or leave the grid at the individual vertices (positive for generators and negative for loads). As shown in Fig.~\ref{fig-capdeg}, the generating capacities of power stations are highly correlated with their degrees, $k_i$. 
Assuming that the whole grid is in power balance, we approximate 
$I_i = \frac{k_ i}{\sum_{\rm generators}k_j}$ for generators and
$I_i =-\frac{k_ i}{\sum_{\rm loads}k_j}$ for loads.

Weighting the two requirements equally (other weighting choices could be made), we attempt to maximize by MC simulated annealing the quality measure 
\be
E=\frac{Q}{Q_{\rm init}} - \sqrt{ \frac{\sum_i I_i^2 }{(\sum_i I_i^2)_{\rm init})}},
\label{eq-E}
\ee
where the subscript ``init'' designates the value {\it after} the first
recombination, but {\it before} any MC steps. The MC steps consist in moving loads that are peripheral to one island to a neighboring island to which it is also connected. 
 
\begin{figure}
\begin{center}
\includegraphics[angle=0,width=0.52\textwidth]{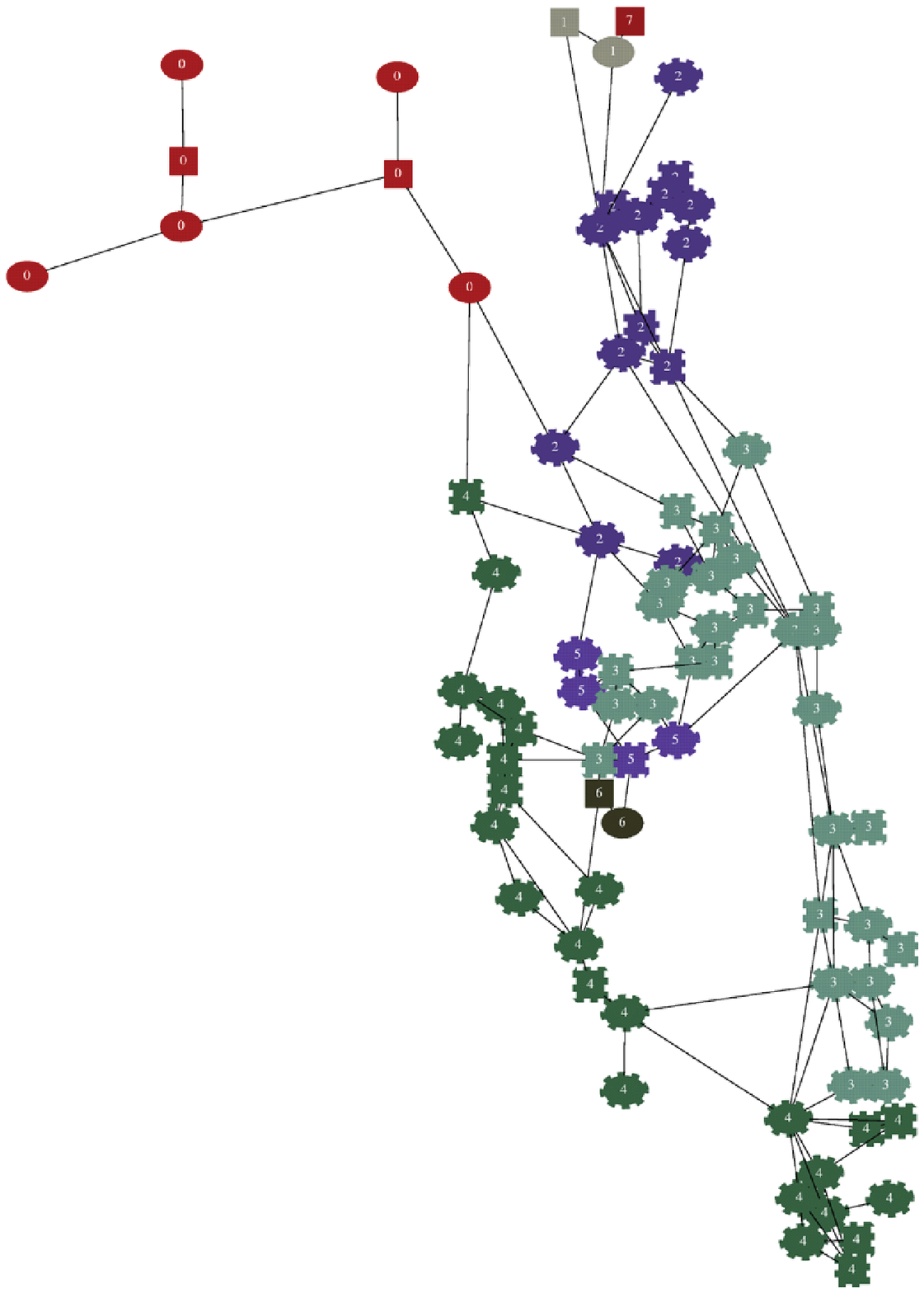} 
\hspace{0.1truecm}
\includegraphics[angle=0,width=0.53\textwidth]{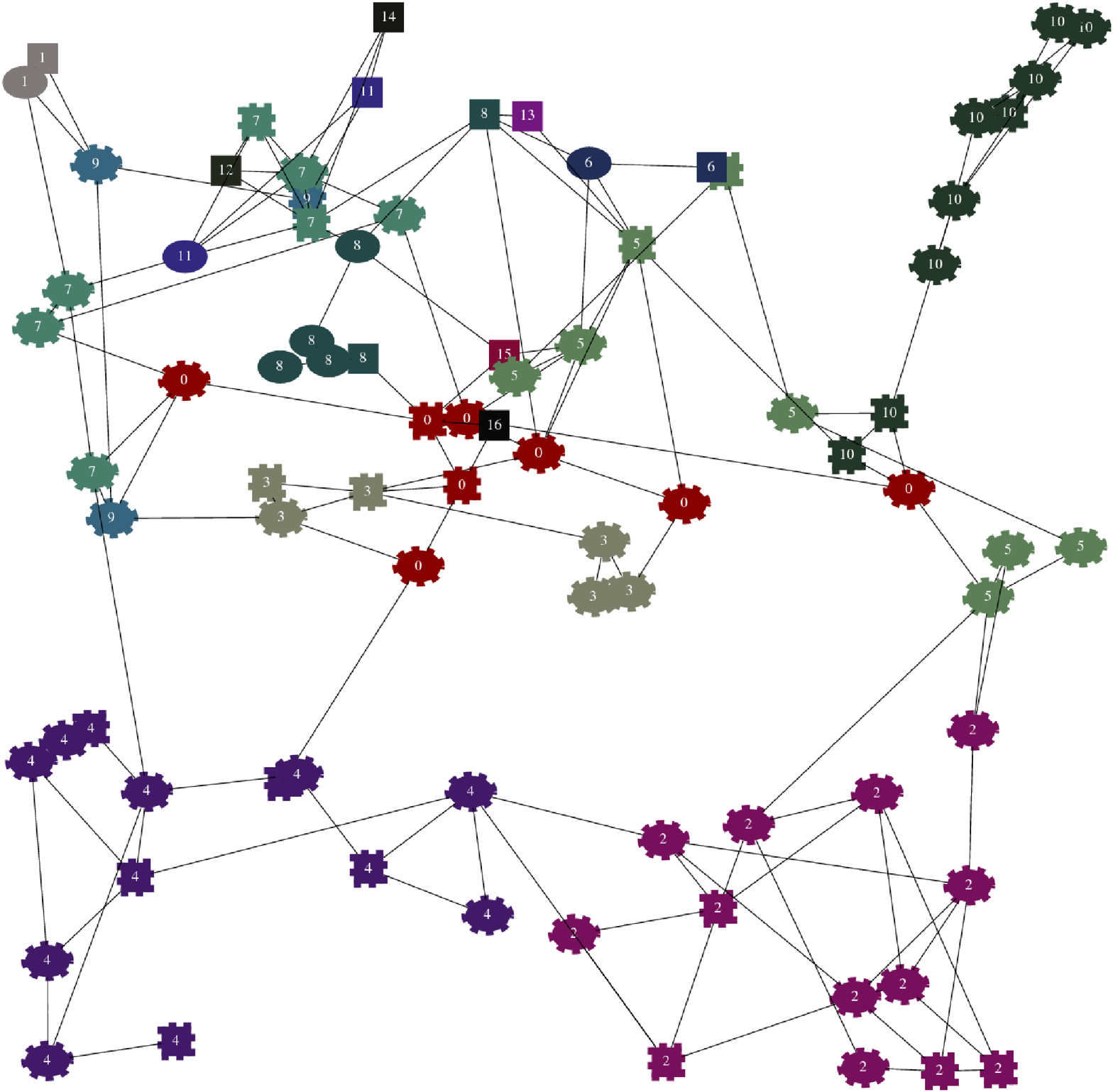} 
\end{center}
\caption[]{
The Florida high-voltage power grid (top) and a representative model network 
(bottom). Generators are represented by squares and loads by ovals. 
The partitions shown are in each case the ``best" ones obtained in ten 
independent runs with the bottom-up partitioning algorithm of 
Ref.~\protect\cite{HAMAD2011}. Islands are identified by different colors 
and by numbers that can be seen if viewed at high magnification. 
See text for discussion.
}
\label{fig-FL}
\end{figure}

\begin{figure}
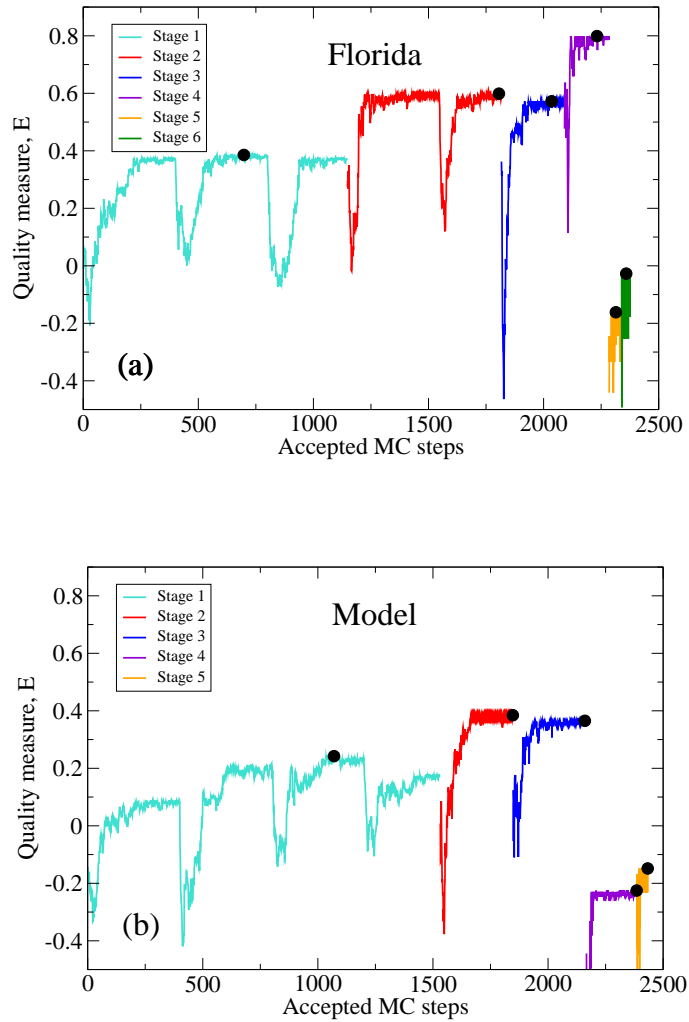

\begin{center}
\includegraphics[angle=0,height=0.50\textwidth]{energyfile_Run5_UGA12.eps} 
\vspace{1.0truecm}
\end{center}
\begin{center}
\includegraphics[angle=0,height=0.50\textwidth]{energyfileRFL6G.eps} 
\end{center}
\caption[]{
Quality measure $E$ vs.\ accepted MC steps in the partitioning algorithm for Florida (a) and our representative model system (b). The black dots represent the maximum value of $E$ obtained in each stage of the partitioning algorithm. The global maxima in both cases represent the highest $E_{\rm max}$ obtained over ten independent partitioning runs. These maxima correspond to the partitionings shown in Fig.~\protect\ref{fig-FL}. 
}
\label{fig-CUR}
\end{figure}

\noindent
{\bf Iteration.}
The optimized islands form a new network (analogous to real-space renormalization-group calculations), in which each
island is represented by a vertex. The connections
between the new vertices are the same as those between the previous
islands. This defines a new conductivity
matrix. The components of the new current vector, $|  \tilde{I} \rangle $ represent the generating surplus or
deficiency of each of the old islands or new ``super-generators'' (smooth symbols in Fig.~\ref{fig-FL}) or
``super-loads'' (``geared" symbols in Fig.~\ref{fig-FL}), respectively. 
This process of agglomeration and optimization is iterated
until all the original vertices belong to one island, and the optimum partitioning is identified. The quality measure $E$
is shown ${\it vs.}$ accepted MC step in
Fig.~\ref{fig-CUR}(a) for the real Florida grid, and for our representative model system in Fig.~\ref{fig-CUR}(b).

\section{Results and Conclusion}
\label{sec:res}
The model power grids introduced here were constructed to match the size, proportion of generators, average degree, and scaled edge length of the real Florida high-voltage grid. In fact, both the degree distribution and the full edge-length distribution for the two networks are quite similar. Nevertheless, it is easier for our partitioning algorithm to find a partition with a high value of $E$ for the real Florida grid than for the models. The resulting partitionings, shown in Fig.~\ref{fig-FL}, consist of eight islands with $E_{\rm max} \approx 0.8$ for Florida and 17 islands with $E_{\rm max} \approx 0.4$ for the representative model. For Florida, the four largest islands comprise 75 of the 84 vertices. The "North-West" portion of the model grid appears particularly difficult to partition. We believe this indicates that the real power grid is  more strongly correlated than the randomized models. 

\section*{Acknowledgments}
\label{sec:ack}
This work was supported in part by 
U.S.\ National Science Foundation Grant No.\ DMR-1104829, 
U.S.\ Office of Naval Research Grant No.\ 
N00014-08-1-0080, and the Institute for Energy Systems, Economics, 
and Sustainability at Florida State University. 







\clearpage

\clearpage

\clearpage


\clearpage

\end{document}